\documentclass{jetpl}
\usepackage{color,amsmath,amsfonts,graphics,epsfig,amssymb}
\usepackage[english]{babel}
\usepackage{xcolor}
\definecolor{xlinkcolor}{cmyk}{1,1,0,0}
\usepackage[bookmarks=true, pdfnewwindow=true, colorlinks=true, linkcolor=xlinkcolor, citecolor=xlinkcolor, filecolor=xlinkcolor, urlcolor=xlinkcolor, final=true]{hyperref}
\twocolumn
\lat

\newcommand{\red}{\color{red}}
\newcommand{\black}{\color{black}}
\renewcommand{\red}{\black}
\newcommand{\redd}{\red}
\title{
\begin{flushright}
\normalsize \rm INR-TH-2024-018
\end{flushright}
\vspace{8mm}
Stellar evolution and axion-like particles: new constraints and hints from globular clusters in the GAIA DR3 data
}

\rtitle{Axions and globular clusters}

\sodtitle{Stellar evolution and axion-like particles: new constraints and hints from globular clusters in the GAIA DR3 data}

\author{S.\,V.\,Troitsky\thanks{email: {\tt st@ms2.inr.ac.ru}}}

\rauthor{S.\,V.\,Troitsky}

\sodauthor{S.\,V.\,Troitsky}

\address{Institute for Nuclear Research of the Russian Academy of
Sciences,\\
60th October Anniversary prospect 7A, 117312 Moscow, Russia\\
and\\
Faculty of Physics, Lomonosov Moscow State University, 1-2 Leninskiye Gory, 119991 Moscow, Russia}

\dates{October 3, 2024}{November 12, 2024; accepted December 11, 2024}

\abstract{
Axion-like particles (ALPs) are hypothetical pseudoscalar bosons, natural in extensions of the Standard Model. Their interactions with ordinary matter and radiation are suppressed, making it challenging to detect them in laboratory experiments. However, these particles, produced within stellar interiors, can provide an additional mechanism for energy loss, potentially influencing stellar evolution. Prominent methods for searching for such effects involve measuring the properties of red giants and helium-burning stars in globular clusters (GCs).
Here we use published catalogs of stars selected as members of seven GCs on the basis of parallaxes and proper motions measured by {\sl Gaia} (Data Release 3). Making use of previously derived theoretical relations and the new data, we find the upper limit on the ALP-electron coupling, $g_{ae}< 1.3 \times 10^{-13}$~(95\%~CL), and an indication ($2.6\sigma$) to nonzero ALP-photon coupling, $g_{a\gamma}=\left( 6.6^{+1.3}_{-1.8}\right) \times 10^{-11}$~GeV$^{-1}$. 
Given the precision of contemporary observational data, it is imperative to refine ALP constraints through more sophisticated analyses, which will be explored in detail elsewhere.
}
\begin{document}
\maketitle

\noindent\textbf{1. Introduction.} 
In many extensions of the Standard Model (SM) of particle physics, axion-like particles (ALPs) are predicted. They are pseudo-Goldstone bosons in two-scale theories, in which a global $U(1)$ symmetry is broken both spontaneously (generating the effective interaction between ALPs and photons, as well as, possibly, with other SM particles) and explicitly (providing a small ALP mass). These two scales may be either related to each other, like in the case of the canonical axion of Quantum Chromodynamics, or be kept as two independent parameters of the model for generic ALPs. Both theory and experiment motivate these particles to interact weakly with ordinary matter and radiation, which makes it hard to detect them experimentally. Their effects can however be observed in astrophysics, see Refs.~\cite{Raffelt:book,DiLuzio2021rev,Raffelt2024rev} for reviews and further references. One well-known approach is to search for the impact of ALPs on the stellar evolution. Thermally produced in central regions of stars, these hypothetical particles would escape freely outside, carrying energy out. In addition to direct searches for such ALPs produced in the Sun, it is possible to consider the effects of the energy losses on the evolution of other stars, observed in their ensembles. 

Of particular interest are the stars at their late stages of evolution in globular star clusters (GCs), where they have similar ages and chemical composition \cite{GCre}. Before helium fusion reactions start in the center, a star passes through the red-giant stage, and extra energy losses could delay the helium ignition. As a result, the brightest red giant may become brighter than expected, shifting the position of the tip of the red-giant branch (TRGB) in the color-magnitude diagram. The subsequent stage of helium burning (HB), when the star moves to the horizontal branch of the same diagram, becomes \red shorter \black in case of higher losses, and the number of stars in this branch, $N_{\rm HB}$, \red decreases \black compared to the standard case. Both observables have been exploited to constrain ALP couplings for decades.

Stellar astrometry and photometry were revolutionized in the last years, when data from {\sl Gaia} \cite{Gaia:mission} started to become available. In early data releases, crowded fields like GCs were not well resolved, but the latest Data Release~3 (DR3) \cite{Gaia:DR3}, together with dedicated GC studies based on it, opens up the possibility to use these high-precision data for constraining ALP physics. This is the aim of the present work.

\vskip 1mm
\noindent\textbf{2. Data.}
In a series of recent papers \cite{gcP1,gcP2,gcP3,gcP4,gcP5,gcP6}, \begin{table*}
\begin{center}
\begin{tabular}{cccccccc}
\hline
 NGC& Ref.& $N_{\mbox{\sl \scriptsize Gaia}}$& $N_{\rm synt}$& distance, kpc       &     [Fe/H]   &$M_{\rm bol}^{\rm (tip)}$  &  $R$\\
\hline
 288& \cite{gcP3}  & 3923 & 439   &$ 8.99  \pm   0.09  $ &  $  -1.3 \pm   0.1 $ & $  -2.770 \pm   0.065$ & $ 1.403 \pm 0.139$\\
 362& \cite{gcP3}  & 4139 & 442   &$ 8.83  \pm   0.10  $ &  $  -1.3 \pm   0.1 $ & $  -3.547 \pm   0.086$ & $ 1.742 \pm 0.185$\\
6218& \cite{gcP3}  & 6231 & 830   &$ 5.11  \pm   0.05  $ &  $  -1.3 \pm   0.1 $ & $  -3.804 \pm   0.102$ & $ 2.900 \pm 0.367$\\
6362& \cite{gcP4}  & 5069 & 582   &$ 7.65  \pm   0.07  $ &  $  -1.04\pm   0.07$ & $  -3.000 \pm   0.067$ & $ 1.446 \pm 0.155$\\
6723& \cite{gcP4}  & 2207 & 373   &$ 8.27  \pm   0.10  $ &  $  -1.09\pm   0.06$ & $  -3.591 \pm   0.076$ & $ 0.905 \pm 0.098$\\
6397& \cite{gcP5}  &17312 &5281   &$ 2.482 \pm   0.019 $ &  $  -1.8 \pm   0.1 $ & $  -3.806 \pm   0.086$ & $ 1.306 \pm 0.138$\\
6809& \cite{gcP5}  & 8828 & 724   &$ 5.348 \pm   0.052 $ &  $  -1.7 \pm   0.1 $ & $  -3.387 \pm   0.098$ & $ 1.035 \pm 0.132$\\
\hline
\end{tabular}
\end{center}
\caption{\label{tab:clusters} \sl
\textbf{Table~\ref{tab:clusters}.} Globular clusters used in this work: NGC names, references, number of stars identified in {\sl Gaia} DR3, number of stars with published {\sl Gaia} synthetic photometry, distance from Ref.~\cite{distances}, metallicity from Refs.~\cite{gcP3,gcP4,gcP5} (uncertainty set to 0.1 if not quoted), and evolution parameters determined here \redd with their cluster-specific uncertainties \black (see Sec.~3).}
\end{table*}Gontcharov et al.\ identified individual members of 14 Galactic GCs, 
making use of parallaxes and proper motions from {\sl Gaia}. For 7 out of these 14 clusters, see Table~\ref{tab:clusters}, Refs.~\cite{gcP3,gcP4,gcP5} published lists of identifiers of these stars in the {\sl Gaia} DR3 database. We use these lists to extract photometry for the cluster member stars from the database, \url{https://www.cosmos.esa.int/web/gaia/dr3}. In order to be consistent with previous studies, we use the synthetic photometry catalog \cite{Gaia:UBVRI} to obtain magnitudes in the {\it UBVRI} colors, available for brighter objects which include all red giants and HB stars in the clusters of interest. We also make use of the iron content [Fe/H] and color corrections $E(B-V)$ 
quoted in \cite{gcP3,gcP4,gcP5}, the extinction correction $A_V$ in the $V$ band from \cite{GALExtin} (model A), and the distances $d$ to GCs from \cite{distances}\red, with their corresponding statistical and systematic uncertainties\black.

\vskip 1mm
\noindent\textbf{3. Analysis.}
We follow Ref.~\cite{StranieroTRGB} in the TRGB analysis and Ref.~\cite{HBstars} in the HB analysis for individual GCs. The combination of data is discussed below in Sec.~3.4.
\vskip 0.5mm
\noindent{\sl 3.1. Determination of the evolution parameters.} 

{\sl (i)~TRGB.} The relevant parameter is the bolometric absolute magnitude of TRGB, $M_{\rm bol}^{\rm (tip)}$, which is determined as follows. One selects the brightest red giant and determines its absolute $V$-band magnitude $V_0=V-\mu-A_V$ from the observed magnitude $V$ and the distance modulus $\mu=5\log_{10}(d/({\mbox{10~pc}}))$. Following Ref.~\cite{StranieroTRGB}, we take the bolometric correction $b$ from Ref.~\cite{bolometric}, where it is tabulated as a function of [Fe/H] and $(V-I)_0 \approx V-I +1.25E(B-V)$ \cite{StranieroTRGB}. Then the absolute bolometric magnitude of the brightest red giant is $M_{\rm bol}^{(0)}=V_0+b$.

It is not a full story because TRGB is not determined by the currently brightest red giant, but instead by the brightest point this red giant can reach in its evolution. This is tackled by introducing a correction $\delta(n)$, determined by simulations in Ref.~\cite{StranieroTRGB} and depending on the number $n$ of red giants with $V$ magnitudes not weaker than $V_0+2.5^{\rm m}$. Finally, one finds $M_{\rm bol}^{\rm (tip)} = M_{\rm bol}^{(0)} - \delta(n)$. 

{\sl (ii)~HB stars.} The relevant parameter is the ratio $R=N_{\rm HB}/N_{\rm RGB}$ of the number $N_{\rm HB}$ of HB stars to the number $N_{\rm RGB}$ of stars in the upper part of the red-giant branch in the color-magnitude diagram for the cluster. $N_{\rm RGB}$ is defined as the number of red giants with the absolute $V$ magnitude brighter than the zero-age HB magnitude $M_{\rm ZAHB}$. The latter magnitude is estimated in Eq.~(1) of Ref.~\cite{ZAHB} for a given metallicity [M/H], which, in turn, is estimated from [Fe/H] using Eq.~(2) of Ref.~\cite{StranieroTRGB}.  One needs to visually separate red giants from asymptotic giants, and to visually identify the horizontal branch, in the color-magnitude diagram.

\vskip 0.5mm
\noindent{\sl 3.2. Relations to the ALP couplings.}

{\sl (i)~TRGB and the electron coupling.} Given the present upper limits on the ALP couplings to various particles, the strongest effect on the evolution of red giants would come from the possible coupling of ALPs $a$ to electrons $e$, which enters the Lagrangian as the Yukawa interaction,
$$
\mathcal{L}_{ae}= g_{ae} a \bar{e} \gamma_5 e.
$$
In Ref.~\cite{StranieroTRGB}, $M_{\rm bol}^{\rm (tip)}$ was obtained from simulations for different values of $g_{ae}$ and [M/H]. As it is a smooth function, we use the interpolation of its values from Fig.~4 in Ref.~\cite{StranieroTRGB}. 

{\sl (ii)~HB stars and the photon coupling.} Neglecting $g_{ae}$ (we will see below that this assumption is justified by the results of our TRGB analysis), the dominant contribution to the energy losses in HB stars would be related to the ALP coupling $g_{a\gamma}$ with two photons,
$$
\mathcal{L}_{a\gamma}=-\frac{1}{4} g_{a\gamma} a F_{\mu\nu} \tilde{F}^{\mu\nu},
$$
where $F_{\mu\nu}$ is the electromagnetic field stress tensor and $\tilde{F}^{\mu\nu}$ its dual. In the particle-physics system of units, the dimension of $g_{a\gamma}$ is 1/mass. We use the relation proposed in Ref.~\cite{HBstars} on the basis of numerical simulations,
\begin{equation}
R = 6.26\,Y-0.41 \left( \frac{g_{a\gamma}}{10^{-10}~\mbox{GeV}^{-1}} \right)^2 -0.12,
\label{Eq:R}    
\end{equation}
where $Y$ is the initial helium fraction of the stellar matter, see below.

\vskip 0.5mm
\noindent{\sl 3.3. Uncertainties.}
We list here the sources of uncertainties in the determination of $R$ and $M_{\rm bol}^{\rm (tip)}$ for individual GCs. Cluster-specific uncertainties are added in quadrature to determine the error bars reported below. Common theoretical uncertainties are treated separately in the combined likelihood, as described in Sec.~3.4. The uncertainty related to $Y$ is discussed below.

{\sl (i)~Statistical uncertainties.}
Both $R$ and $M_{\rm bol}^{\rm (tip)}$ suffer from availability of only a finite number of stars in a GC. To evaluate the corresponding statistical uncertainty for $R$, we assume that $N_{\rm HB}$ and $N_{\rm RGB}$ follow the Poisson statistics. For $M_{\rm bol}^{\rm (tip)}$, this effect is accounted by uncertainties in $\delta(n)$ presented in Ref.~\cite{StranieroTRGB}.

{\sl (ii)~Parameter uncertainties.}
Uncertainties in the distances, metallicities and color corrections were reported together with their values. Here we assume that they follow Gaussian distributions.  \red We use the uncertainties estimated in Ref.~\cite{StranieroTRGB} for the bolometric corrections.\black

{\sl (iii)~Theoretical uncertainties.}
The present study uses simplified numerical relations from Refs.~\cite{StranieroTRGB,HBstars}, based on numerical simulations performed within certain assumptions. The corresponding theoretical uncertainty in $M_{\rm bol}^{\rm (tip)}$ was estimated in Ref.~\cite{StranieroTRGB} \redd as 0.038$^{\rm m}$\black; we \redd however use a more conservative estimate of 0.12$^{\rm m}$ from Ref.~\cite{TRGB-theory} \black  here. \redd The latter estimate accounts for possible variations of the mass of the stars leaving the main sequence now, related to their helium content. \black

Many uncertainties cancel in the $R$ ratio, and theoretical uncertainties of $R$ are subdominant \cite{DiLuzio:Xe,NewNuclearRatesEffect}.
\red 
The main theoretical uncertainty is related to the numerical description of convection. It was studied in Ref.~\cite{R2}; we estimate \redd this systematic uncertainty \black as 4.5\% in $R$ and take into account in the analysis.
\black

{\sl (iv)~Helium abundance.}
As can be seen from Eq.~(\ref{Eq:R}), $R$ depends strongly on the helium abundance $Y$, which is hard to measure in particular sources. Since GCs are old objects without late-time stellar formation \cite{GCre}, it is often assumed that the helium abundance there is close to primordial, $Y_{\rm BBN}=0.245 \pm 0.003$ \cite{PDG}. In very few cases, $Y$ was determined spectroscopically in GCs. Notably this includes NGC~6397, the nearest of the seven clusters we use here, for which one has $Y=0.241 \pm 0.004$ \cite{Y-NGC6397}, in a perfect agreement with $Y_{\rm BBN}$. Primordial values of $Y$ are conservative for ALP searches, because the same $R$ would require larger $g_{a\gamma}$ for larger $Y$. We report the results for $Y=Y_{\rm BBN}$ as fiducial ones and show how they are changed with $Y$ in Sec.~4.

\vskip 0.5mm
\noindent{\sl 3.4. Combination of measurements.}
To account for the ensemble of measurements for different GCs, which have different metallicities, we follow the Bayesian approach, using the conditions $g_{ae} \ge 0$ and $g_{a\gamma} \ge 0$ as priors. Let $g$ be one of these coupling constants and denote as $x$ the corresponding evolution parameter, measured as $x_i \pm \delta x_i$ in the $i$-th GC. The theoretical value of the observable for this cluster is given by a function $x_{\rm th}(k_i,g)$ of $g$ and of the value $k_i$ of a cluster parameter. Namely, for $g=g_{ae}$, $x=M_{\rm bol}^{\rm (tip)}$, $k_i=$[M/H]$_i$, and the function $x_{\rm th}(k_i,g) \equiv M_{\rm bol}^{\rm (tip)} (\mbox{[M/H]}_i, g_{ae})$ is taken from Ref.~\cite{StranieroTRGB}, see Sec.~3.2(i). For $g=g_{a\gamma}$, $x=R$, $k_i=Y$ and $x_{\rm th}(k_i,g) \equiv R(Y,g_{a\gamma})$ is given by Eq.~(\ref{Eq:R}).

Since the dominant theoretical uncertainties are common to all clusters, they must be treated as correlated rather than included independently in the uncertainties of individual measurements.
We introduce the likelihood function,
\begin{equation}
L(g)=\theta(g)\int_{-\infty}^{\infty}
\frac{dt}{\sqrt{2\pi}}\,e^{-t^2/2}
\prod_i P(x_i|g,t),
\label{Eq:likelihood}
\end{equation}
where $\theta(g)$ is the Heaviside step function, $t$ is a common Gaussian nuisance parameter describing the correlated theoretical uncertainty, 
\begin{equation}
P(x_i|g,t)=
\frac{1}{\sqrt{2\pi}\,\delta x_i}
\exp\left[-\frac{r_i^2(g,t)}{2(\delta x_i)^2}\right],
\end{equation}
and $r_i(g,t)=x_i-x_{\rm th}(k_i,g)-t\,\sigma_{{\rm th},i}(g)$. Here $\delta x_i$ includes the uncertainties uncorrelated between clusters, while $\sigma_{{\rm th},i}(g)$ describes the common theoretical uncertainty, $\sigma_{{\rm th},i}=0.12^{\rm m}$ for TRGB and $\sigma_{{\rm th},i}=0.045\,x_{\rm th}(k_i,g)$ for $R$.
The integral over $t$ is evaluated analytically.
Once $L(g)$ is constructed, the best-fit value of $g$ corresponds to the maximal value of $L$. The confidence interval for $g$  at the confidence level $\xi$ is determined by the condition $L(g)>L_0$ such that 
\begin{equation}
\int\limits_{L(g)>L_0} \!\!\!\! L(g)\, dg = \xi \int\limits_{-\infty}^{\infty}\! L(g)\, dg.
\label{Eq:L-CL}    
\end{equation}

\begin{figure}
    \centering
    \includegraphics[width=\linewidth]{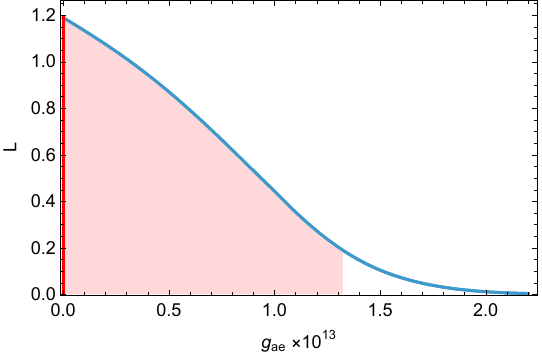}
    \caption{\label{fig:Lg_ae} \sl
\textbf{Figure~\ref{fig:Lg_ae}.}
Normalized likelihood $L$ profile for $g_{ae}$ obtained in the present study. $L$ is maximal at $g_{ae}=0$, and the shaded region presents the 95\%~CL range of allowed couplings.
}
\end{figure}

\begin{figure}
    \centering
    \includegraphics[width=\linewidth]{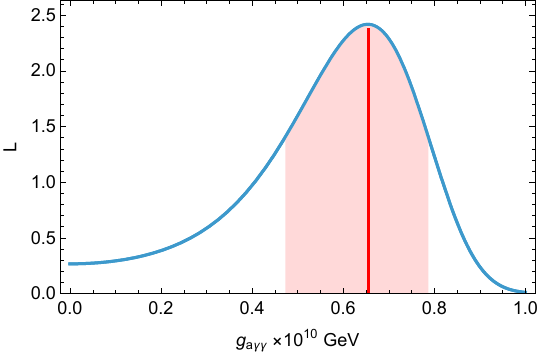}
    \caption{\label{fig:Lg_ag} \sl
\textbf{Figure~\ref{fig:Lg_ag}.}
Normalized likelihood $L$ profile for $g_{a\gamma}$ obtained in the present study for $Y=Y_{\rm BBN}$. The value of $g_{a\gamma}$ which maximizes $L$ is shown by the vertical line, and the shaded region presents the 68\%~CL range of allowed couplings.
}
\end{figure}

\begin{figure}
    \centering
    \includegraphics[width=\linewidth]{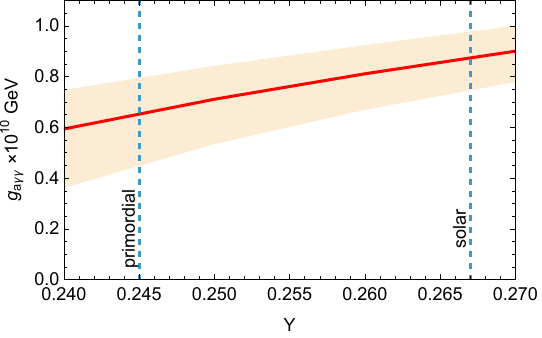}
    \caption{\label{fig:Y} \sl
\textbf{Figure~\ref{fig:Y}.}
Best-fit (full line) and 68\%~CL favored range (shaded region) for the ALP-photon coupling $g_{a\gamma}$ for different assumptions about the helium abundance $Y$. Vertical dashed lines indicate primordial and solar values of $Y$.
}
\end{figure}

\vskip 1mm
\noindent\textbf{4. Results.}
Values of $M_{\rm bol}^{\rm (tip)}$ and $R$ obtained here are presented in Table~\ref{tab:clusters} together with other parameters of seven individual clusters. Similarly to previous studies, cf.\ e.g.\ Table~2 in Ref.~\cite{prevRmeasurements}, they demonstrate considerable scatter \redd which may be studied in future with larger number of GCs\black, and the main results of the present work come from the likelihood analysis described in Sec.~3.4. Figures~\ref{fig:Lg_ae} and \ref{fig:Lg_ag} present the resulting $L$ profiles for $g_{ae}$ and $g_{a\gamma}$, respectively. 
Our analysis favors $g_{ae}=0$ as the best-fit value and sets the upper limit of $g_{ae}< \redd 1.3 \black \times 10^{-13}$~(95\%~CL), while, in the most conservative assumption of $Y=Y_{\rm BBN}$, nonzero $g_{a\gamma}=\left( 6.6^{+1.3}_{-1.8}\right) \times 10^{-11}$~GeV$^{-1}$ (68\%~CL uncertainty) is preferred. The confidence level, at which $g_{a\gamma}=0$ is disfavored at $Y=Y_{\rm BBN}$, can be determined by putting $L_0=L(0^+)$ in Eq.~(\ref{Eq:L-CL}), giving $1-\xi=0.0098$, which would correspond to a $2.6\sigma$ indication for the Gaussian statistics.
Results for larger helium abundances are presented in Fig.~\ref{fig:Y}.
Like other results related to stellar energy losses, those obtained in this work do not depend on the ALP mass $m$ provided it is much smaller than the temperature in the stellar interiors ($\sim$keV).

\vskip 1mm
\noindent\textbf{5. Discussion.}
ALP interactions with ordinary matter may be studied in a plethora of approaches, see e.g.\ Ref.~\cite{ALPexp-rev} for a review. They may be divided in four groups.

{\sl Laboratory experiments.} They usually provide the most robust, though weak, constraints on the couplings.

{\sl Laboratory detection of astrophysical ALPs.} These include the experimental search for ALPs produced in the Sun by means of well-understood processes in the solar central region, as well as various direct searches for ALPs as dark-matter particles (in the assumption that the dark matter consists of ALPs, which does not hold in general). 

{\sl Astrophysical searches not relying on magnetic-field models.} These are dominated by studies of stellar energy losses at various stages of evolution, including supernova explosions. They are based on assumptions about processes in stellar interiors, which are qualitatively robust but still allowing for quantitative model dependence. The present study falls in this group.

{\sl Searches for ALP-photon conversion in astrophysical magnetic ($B$) fields.} This conversion may manifest itself in various features in high-energy spectra of astrophysical objects, including suppression or lack thereof, irregularities etc. These constraints are often the strongest among the four groups, but they depend on the assumed values and configurations of poorly known cosmic magnetic fields, and the corresponding uncertainties may be large, see e.g.\ Refs.~\cite{LT-magnetic-fields,k-Mrk}.

To put our results in context, we compare them (see Table~\ref{tab:electron} for $g_{ae}$ and Table~\ref{tab:gamma} for $g_{a\gamma}$) with the strongest previously published limits from each of the four groups. A reader interested in a wider landscape of limits is directed to Refs.~\cite{PDG,AxionLimits}, where dozens of other constraints are reported. For definiteness, we fix $m=10^{-8}$~eV for a few cases when the results are mass-dependent. 
\begin{table}[]
    \centering
    \begin{tabular}{cccc}
         \hline
\multicolumn{4}{c}{ALP-electron coupling}\\         
         \hline
group    & experiment & Ref. & $g_{ae}$, \\
& (technique) & & $10^{-13}$\\
\hline
laboratory          & torsion&\cite{TorsionPendulum} & $<74500$ \\
                    & pendulum  &  & \\
solar               & XENONnT & \cite{XENONnT} & $<19$ \\
stellar             & TRGB & \cite{StranieroTRGB} & $<1.48$ \\
\hline
\multicolumn{3}{c}{this work} & $<1.3$ \\
               \hline
    \end{tabular}
    \caption{\label{tab:electron} \sl
\textbf{Table~\ref{tab:electron}.} Strongest constraints on $g_{ae}$ obtained  by various methods. $g_{a\gamma}=0$ and $m=10^{-8}$~eV are assumed. The XENONnT constraint is at 90\%~CL, others are at 95\%~CL.}
    \end{table}
\begin{table}[]
    \centering
    \begin{tabular}{cccc}
         \hline
\multicolumn{4}{c}{ALP-photon coupling}\\         
         \hline
group    & experiment & Ref. & $g_{a\gamma}$, \\
& (technique) & & $10^{-11}$~GeV$^{-1}$\\
\hline
laboratory          & OSQAR&\cite{OSQAR} & $<3550$ \\
solar               & CAST & \cite{CAST,CAST2024} & $<5.7$ \\
stellar             & AGB stars & \cite{R2} & $<4.7$ \\
$B$ field           & pulsars   & \cite{PSR}& $<0.4$\\
    \hline
\multicolumn{3}{c}{this work} & $6.6^{+1.3}_{-1.8}$ \\
            \hline
    \end{tabular}
    \caption{\label{tab:gamma} \sl
\textbf{Table~\ref{tab:gamma}.} Strongest constraints (95\%~CL for upper limits) on $g_{a\gamma}$ obtained  by various methods. $g_{ae}=0$ and $m=10^{-8}$~eV are assumed.}
    \end{table}

The comparison suggests that our upper limit on $g_{ae}$ is stronger than previously reported ones. This may be attributed to more effective selection of GC members with {\sl Gaia} data, which reduces the contribution of non-GC stars projected to the GC direction. Our constraint is in marginal tension with indications to nonzero $g_{ae}=1.6^{+0.29}_{-0.34}\times 10^{-13}$ from white-dwarf cooling \cite{StellarRevAndWDhint}.

Nonzero $g_{a\gamma}$ preferred by our results agrees with previous laboratory and, marginally, solar and stellar limits. Notably, they are consistent with previous studies which used the $R$ parameter and gave weak indications to $g_{a\gamma}>0$. In particular, Ref.~\cite{HBstars} found the best-fit $g_{a\gamma}=\left( 4.5^{+1.2}_{-1.6}\right) \times 10^{-11}$~GeV$^{-1}$ and put the 95\%~CL upper limit of $g_{a\gamma}< 6.6\times 10^{-11}$~GeV$^{-1}$. In Ref.~\cite{R2}, a stronger upper limit, see Table~\ref{tab:gamma}, was obtained from studies of asymptotic giant branch (AGB) stars, while the HB stars $R$ parameter again slightly preferred nonzero $g_{a\gamma} \sim (4-7)\times 10^{-11}$~GeV$^{-1}$. However, some previously published astrophysical studies based on assumptions of values and spatial structure of cosmic magnetic fields claimed stronger upper limits, so that our results are in tension with these model-dependent constraints. 

Some observations of gamma-ray sources at very high or ultra-high energies suggest that the Universe may be more transparent than expected, and this may require new physics (see Refs.~\cite{ST-mini-rev,Roncadelli-review2022} for reviews and further references and e.g.\ \cite{GRB-Galanti,GRB-ST1,GRB-ST2,newHiRes} for more recent advances). This may find its explanation in conversion of an energetic photon to ALP in the magnetic field close to the source and reconversion back to photon close to the observer \cite{Serpico,FRT:2009}. The values of $g_{a\gamma}$ favored for this explanation, $(4-9)\times 10^{-11}$~GeV$^{-1}$, match well the indications obtained in the present work.

\red
The increased precision we find here with respect to previous studies traces back to advantages of {\sl Gaia} DR3 data and overweights the small statistics of the cluster sample. These advantages include the roughly doubled precision in parallax distances of DR3 with respect to various distance indicators used previously and the reduction of the number of field stars in the color-magnitude diagram. \black
Given the dramatic increase in the precision of GC astronomy, it is mandatory to improve stellar-evolution simulations, which are behind this study and now dominate the error budget. Direct simulations of observable quantities, taking into account more potential variations of stellar models (including description of convection), should be applied to a larger number of {\sl Gaia} GCs. \red End-to-end simulations in terms of {\sl Gaia} magnitudes would further reduce uncertainties related to intermediate steps of the analysis.
\black
These avenues will be followed elsewhere.

While interpreted in terms of ALP couplings, the present study actually constrained only the anomalous energy losses at certain stages of stellar evolution. Other physics may cause these losses, including other hypothetical particles, neutrino properties, etc. These scenarios may be constrained in a similar way to ALPs from  {\sl Gaia} GC data. 

\vskip 1mm
\noindent\textbf{6. Conclusions.}
We searched for non-standard cooling of red giants and helium-burning stars in seven Galactic globular clusters, based on member selection of Refs.~\cite{gcP3,gcP4,gcP5} from {\sl Gaia} DR3 data. We did not find indications for this extra cooling in red giants and interpreted it as the upper limit on the ALP coupling to electrons, the strongest one to date. Contrary, cooling of the helium-burning stars disfavors the zero ALP--photon coupling at the $2.6\sigma$ level. The preferred range of this coupling matches previous hints from stellar evolution and from transparency of the Universe for gamma rays, but is in tension with some model-dependent astrophysical bounds based on the assumptions about cosmic magnetic fields. The returning positive hint for $g_{a\gamma}\ne 0$ motivates further, more detailed studies, which are in progress.

\vskip 1mm
\noindent\textbf{Acknowledgements.} 
The author is indebted to K.~Postnov \redd and to the anonymous reviewer \black for interesting discussions of details of the stellar evolution \redd and of corresponding systematic uncertainties\black, and to G.~Rubtsov for illuminating explanation of the Bayesian approach in the context of the present work. This work was initiated, and essentially performed, during the ``Quarks-2024'' International seminar in Pereslavl', and the atmosphere of the seminar, as well as discussions with participants, are greatly acknowledged.

This work has made use of public data from the European Space Agency (ESA) mission {\sl Gaia} (\url{https://www.cosmos.esa.int/gaia}), processed by the {\sl Gaia} Data Processing and Analysis Consortium (DPAC, \url{https://www.cosmos.esa.int/web/gaia/dpac/consortium}). Funding for the DPAC has been provided by national institutions, in particular the institutions participating in the {\it Gaia} Multilateral Agreement.

\vskip 1mm
\noindent\textbf{Funding.} This work was supported by the Russian Science Foundation, grant 22-12-00253.
\vskip 1mm
\noindent\textbf{Conflict of interest.} The author claims no conflict of interest.

\noindent\textbf{Note added (November 2026).}
The original published version of this paper treated theoretical uncertainties common to different globular clusters as independent. This has been corrected in the present version, resulting in revised numerical constraints and likelihood profiles. 
\bibliographystyle{nature1}
\bibliography{globular}
\end{document}